\documentclass[]{beilstein}

\usepackage{xspace}

\usepackage{amssymb,amsbsy,amsmath}
\newcommand{\fmref}[1]{(\protect\ref{#1})}
\renewcommand{\eqref}[1]{Eq.~(\protect\ref{#1})}
\newcommand{\figref}[1]{Fig.~\protect\ref{#1}}

\begin{document}

%% The title of the article is given with the usual \title command.
%%
%% Each author should be given as a separate \author command.
%%
%% For corresponding authors please use \author* and give the email
%% address as a second mandatory argument.
%%
%% The affiliation of authors is given after the authors; the
%% affiliations are numbered consecutively.
%%
%% If some authors have the same affiliation you can use the optional
%% argument of \author and \author* to give the number of that
%% The whole block is printed with the \maketitle command at the very
%% end.
%%%%%%%%%%%%%%%%%%%%%%%%%%%%%%%%%%%%%%%%%%%%%%%%%%%%%%%%%%%%%%%%%%%%%
\title{Young's moduli of carbon materials investigated by various classical molecular dynamics schemes}
\author{Florian Gayk}
\author{Julian Ehrens}
\author{Tjark Heitmann}
\author{Patrick Vorndamme}
\author{Andreas Mrugalla}
\author*{J\"urgen Schnack}{jschnack@uni-bielefeld.de}
\affiliation{Fakult\"at f\"ur Physik, Universit\"at Bielefeld, Postfach 100131, D-33501 Bielefeld, Germany}
\maketitle

%%%%%%%%%%%%%%%%%%%%%%%%%%%%%%%%%%%%%%%%%%%%%%%%%%%%%%%%%%%%%%%%%%%%%
%% The document should begin with an abstract, if appropriate. If one
%% is given and should not be, a warning is issued.
%%
%% For the three parts of the abstract, ``Background'', ``Results''
%% and ``Conclusions'', the corresponding commands should be used.
%%%%%%%%%%%%%%%%%%%%%%%%%%%%%%%%%%%%%%%%%%%%%%%%%%%%%%%%%%%%%%%%%%%%%
\begin{abstract}
\background Classical carbon potentials together with classical
molecular dynamics are employed to
calculate structures and physical properties of such
carbon-based materials where quantum mechanical methods fail
either due to the excessive size, irregular
structure or long-time dynamics. Examples are given by recently
synthesized 
free-standing carbon nanomembranes (CNM) with molecular
thickness and macroscopic lateral size as well as by amorphous
carbon. 
\results Although such potentials, as for instance implemented in
LAMMPS, yield reasonably accurate bond lengths and angles for
several carbon materials such as graphene, it is not clear how
accurate they are in terms of mechanical properties such as
Young's moduli. We performed large-scale classical molecular
dynamics investigations of three carbon-based materials using 
the various potentials implemented in LAMMPS as well as the
highly sophisticated EDIP potential of Nigel Marks. We demonstrate
how the Young's moduli vary with classical potentials and
compare to experimental results. 
\conclusion Since classical descriptions of carbon are bound to
be approximations it is not astonishing that different
realizations yield differing results. One should therefore
carefully check for which observables a certain potential is
suited. We hope to contribute to such a clarification.
\end{abstract}

%%%%%%%%%%%%%%%%%%%%%%%%%%%%%%%%%%%%%%%%%%%%%%%%%%%%%%%%%%%%%%%%%%%%%
%% Keywords can be given with the \keywords command which takes five
%% arguments. The arguments have to be sorted.
%%%%%%%%%%%%%%%%%%%%%%%%%%%%%%%%%%%%%%%%%%%%%%%%%%%%%%%%%%%%%%%%%%%%%
\keywords{carbon materials; classical molecular dynamics;
  Young's moduli}

%%%%%%%%%%%%%%%%%%%%%%%%%%%%%%%%%%%%%%%%%%%%%%%%%%%%%%%%%%%%%%%%%%%%%
\section{Introduction}
\label{sec-1}

Several carbon-based materials cannot be simulated by quantum
mechanical means, not even by Density Functional Theory (DFT),
since they are either too extended or not regular. The latter is
probably the case for the material we are interested in in
the long run: nanometer thin carbon membranes of macroscopic
lateral size, which are produced from molecular precursors
\cite{GSE:APL99,TBN:AM09,AVW:ASCN13,TuG:AM16}. Although the
precursors are well-characterized, not much is known about the
internal structure of such nanomembranes \cite{MrS:BN14}. Very
likely the material is disordered such as a glass. Mechanical
properties on the other hand, such as Young's moduli, can be
determined \cite{ZBG:B11}. 

Our medium term goal therefore consists in evaluating possible
structures of various carbon nanomembranes by employing classical
molecular dynamics calculations and relating them to mechanical
observables \cite{MrS:BN14}. But since the classical calculations
suffer from their approximate nature, we first want to quantify
their accuracy for Young's moduli for known systems, before
we evaluate moduli for unknown systems. We suspect that the various
potentials that have been developed to date might result in
various structures and various moduli depending on
the employed classical potentials. A very valuable comparison
along these lines, in which the graphitization of amorphous
carbon was studied, has been published recently \cite{TSM:C16}.
As expected, none of the classical potentials works perfectly
for a complex process such as graphitization, and some
potentials perform poorly. For the expert this might guide
future developments, for the user this is a valuable information
on which potential to choose for certain investigations. 

Since the quality of a classical description might very much
depend on the investigated observable, we are continuing the
efforts of \cite{MCM:PRB02,LXS:ASS13,TSM:C16} by investigating the Young's moduli
of three well-known carbon materials in large scale
calculations. As materials we choose graphene, a carbon
nanotube, and diamond. For the simulations we used various
carbon 
interatomic potentials as included in LAMMPS \cite{Pli:JCP95} as
well as the modified EDIP potential of Nigel Marks
\cite{Mar:PRB00} which has been 
demonstrated to be able to simulate extended carbon structures
\cite{MCM:PRB02,PML:PRB09}. 

The article is organized as follows. In the next section we
shortly repeat the essentials of classical molecular 
dynamics calculations. The main section is devoted to the
simulations of the three carbon materials. The article closes
with a discussion.

%%%%%%%%%%%%%%%%%%%%%%%%%%%%%%%%%%%%%%%%%%%%%%%%%%%%%%%%%%%%%%%%%%%%%
\section{Classical carbon-carbon interaction}
\label{sec-2}

A realistic classical carbon-carbon interaction must be able to
account for the various $sp^n$--binding modes. The program
package LAMMPS \cite{Pli:JCP95} offers several of such
potentials, among them those developed by Tersoff and Brenner in
various versions \cite{Ter:PRB88,Bre:PRB90,BSH:JPCM02} as well
as new extensions built on the original potentials.

In addition to the implemented potentials we are going to use
the improved EDIP potential by 
Marks \cite{Mar:PRB00} which so far is not included in standard
versions of LAMMPS. Taking this potential as an example, we want
to qualitatively explain how such potentials work. These
potentials comprise density-dependent two- and three-body
potentials, $U_2$ and $U_3$ in this example respectively,  
%===================    eqn   =================================
\begin{equation}
\label{E-2-1}
U\left(\vec{R}_1,\dots,\vec{R}_N\right) = \sum_{i=1}^{N} \left(
\sum_{\substack{j=1\\j\neq i}}^{N} U_2(R_{ij}, Z(i)) +
\sum_{\substack{j=1\\j\neq i}}^{N} \sum_{\substack{k=j+1\\k\neq
    i}}^{N} U_3(R_{ij},R_{ik},\theta(i,j,k),
Z(i))\right)
\end{equation}
%===================    eqn   =================================
which account for the various binding modes. This is achieved by
an advanced parameterization in terms of a smooth coordination
variable $Z(i)$ as well as by appropriate angle dependencies
$\theta(i,j,k)$. The EDIP potential employs a cutoff of
3.2~\AA\ and a dihedral penalty.

Ground states are then found by the method of steepest
descent, by conjugate gradients or damped dynamics (frictional
cooling). The Young's modulus $E$ in 
the ground state, i.e. at temperature $T=0$~K, 
can be evaluated from the curvature of $U$ at the ground state
configuration (the kinetic energy is zero) \cite{HGB:PRL98}
%===================    eqn   =================================
\begin{equation}
\label{E-2-2}
E_V = \frac{1}{V_0} \left(\frac{\partial^2 U}{\partial \alpha^2}\right)_{\alpha=1}
\ ,
\end{equation}
%===================    eqn   =================================
where $\alpha$ is the dimensionless scaling factor of all
positions and $V_0$ the cuboidic volume of the sample in
equilibrium. For two-dimensional systems such as graphene, which
do not have a volume in classical molecular dynamics, 
\eqref{E-2-2} can be replaced by 
%===================    eqn   =================================
\begin{equation}
\label{E-2-3}
E_S = \frac{1}{S_0} \left(\frac{\partial^2 U}{\partial \alpha^2}\right)_{\alpha=1}
\ ,
\end{equation}
%===================    eqn   =================================
where $S_0$ is the area of the stretched material in
equilibrium \cite{HGB:PRL98}. Several authors introduced an artificial thickness
$h_0$ in order to stay with definition \fmref{E-2-2}. This
thickness is often taken either as the graphite interlayer
distance $h_0=3.35$~\AA\ or the carbon-carbon distance of
graphene, i.e. $h_0=1.42$~\AA. In this article we choose
$h_0=3.35$~\AA.

%%%%%%%%%%%%%%%%%%%%%%%%%%%%%%%%%%%%%%%%%%%%%%%%%%%%%%%%%%%%%%%%%%%%%
\section{Theoretical Investigations}
\label{sec-3}

We included the following carbon potentials in our
investigations: Tersoff in various versions \cite{Ter:PRB88,Ter:PRB89,Ter:PRL90,Ter:PRB94}, 
REBO-II \cite{BSH:JPCM02} and AIREBO as
well as  ABOP \cite{ZWF:JCC15}. The AIREBO potential \cite{STH:JCP00,KRS:JCP03}
is investigated with its
flavours: ``naked" AIREBO, AIREBO with additional long range Lennard-Jones
potential (AIREBO+LJ), AIREBO with additional torsion term
(AIREBO+t), and AIREBO with both terms (AIREBO+LJ+t).
If not otherwise stated, the cutoff of the Lennard-Jones
potential has been chosen as 10.2~\AA.
All of these potentials are discussed in great depth in
Ref.~\cite{TSM:C16}. In addition we performed simulations with
the EDIP potential of Nigel Marks \cite{Mar:PRB00}. For all
potentials the respective ground states are determined, which do
not need to be the same. Then the moduli are evaluated for $T=0$~K.

%%%%%%%%%%%%%%%%%%%%%%%%%%%%%%%%%%%%%%%%%%%%%%%%%%%%%%%%%%%%%%%%%%%%%
\subsection{Graphene}
\label{sec-3-1}

Our theoretical investigations consist in the generation of
initial arrangements of $\sqrt{N} \times\sqrt{N}$ large graphene
sheets with open boundary conditions. As we let $N$ grow to
large numbers, finite size as well as boundary effects
decrease. 

%===================    figure   =================================
%\begin{figure}[ht!]
\begin{figure}
\centering
\includegraphics*[clip,width=80mm,keepaspectratio]{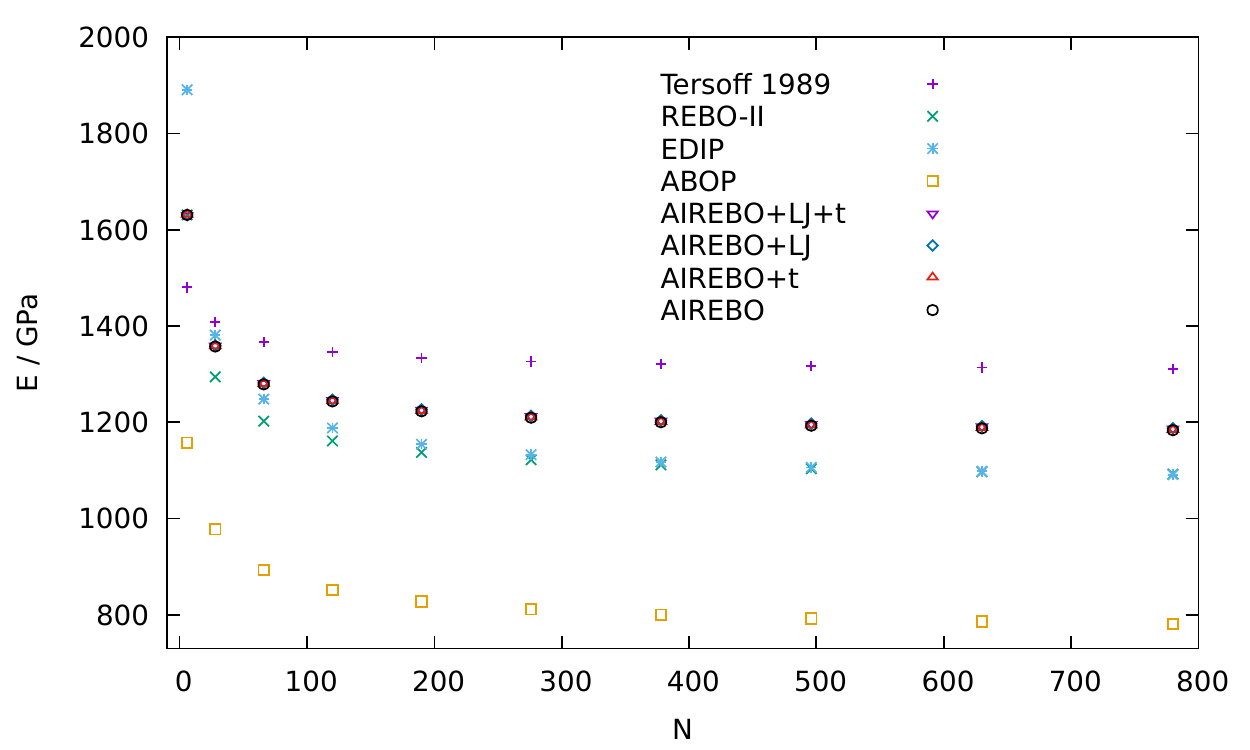}
\includegraphics*[clip,width=80mm,keepaspectratio]{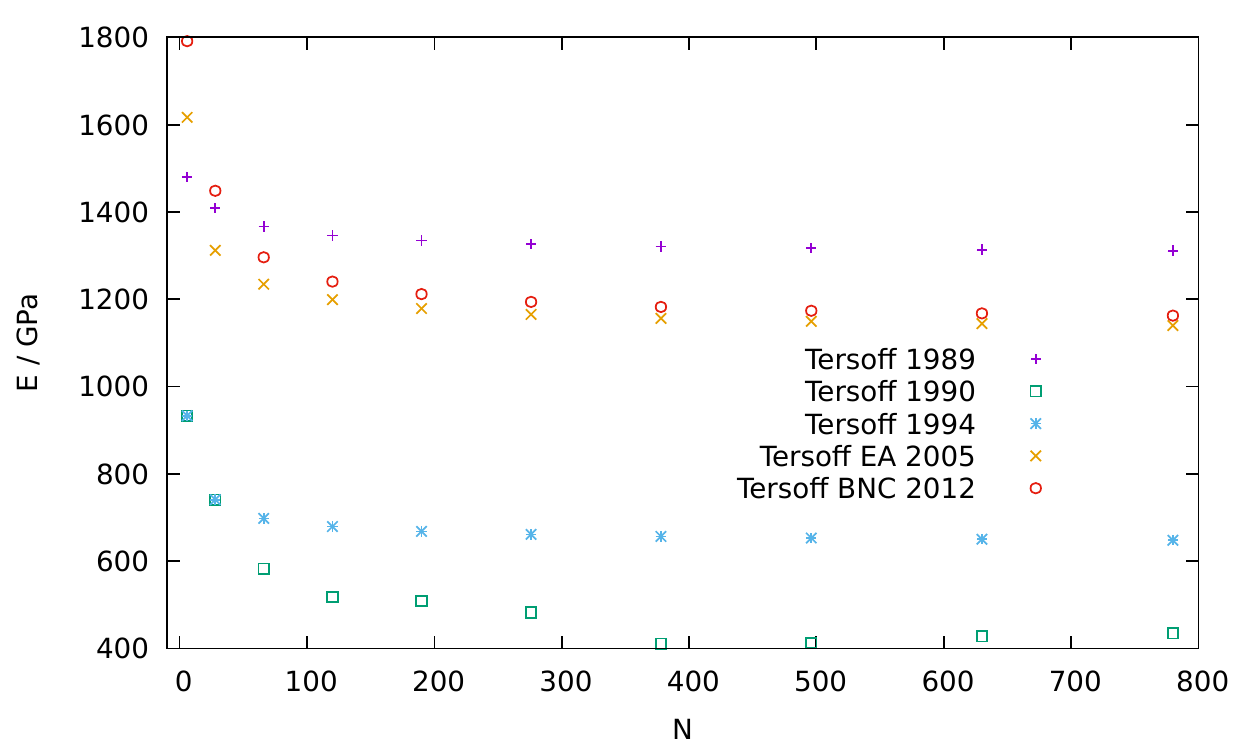}
\caption{Young's modulus of graphene for various sizes and
  potentials.} 
\label{cmd-modulus-graphene}
\end{figure}
%===================    figure =================================

The experimental value for the Young's modulus of graphene is
about 1~TPa \cite{LWK:S08}, which is also reproduced as 1.05~TPa
by DFT calculations for this regular structure \cite{FPJ:PRB07}.
Figure~\ref{cmd-modulus-graphene} shows the results
obtained with the various potentials on the l.h.s., whereas the
r.h.s. displays the moduli obtained for several versions of the
Tersoff potential. The modulus turns out to be isotropic in
accordance with Refs.~\cite{BeB:13,Cao:P14}.
The majority of potentials converges with $N$
against values for the modulus in the range of $1.1\dots
1.3$~TPa. The various investigated AIREBO potentials yield
identical results. The EDIP potential comes closest to
1~TPa, practically on top with REBO-II, 
whereas the ABOP modulus falls below 0.8~TPa.

The chosen Tersoff potentials, displayed on the r.h.s. of
\figref{cmd-modulus-graphene}, exhibit a similar spread of
results. Earlier parameterizations of 1989 and 1994 deviate by
about 0.3~TPa from the value of 1~TPa, whereas the more recent
parameterizations of 2005 and 2012 yield values of 1.1~TPa
similar to the EDIP or REBO-II potentials. It should be noted
that the Tersoff potential of 
1990  does not reproduce the correct graphene structure in
our simulations. 

For C-C- bond distances compare table~\ref{tab-cmd-modulus-1}.

%%%%%%%%%%%%%%%%%%%%%%%%%%%%%%%%%%%%%%%%%%%%%%%%%%%%%%%%%%%%%%%%%%%%%
\subsection{Carbon Nanotubes}
\label{sec-3-2}

The investigated carbon nanotube (CNT) is a (20,20) tube with
armchair geometry. In the investigation we varied the number of
carbon atoms $N$, which is also a measure of length.

%===================    figure   =================================
%\begin{figure}[ht!]
\begin{figure}
\centering
\includegraphics*[clip,width=80mm,keepaspectratio]{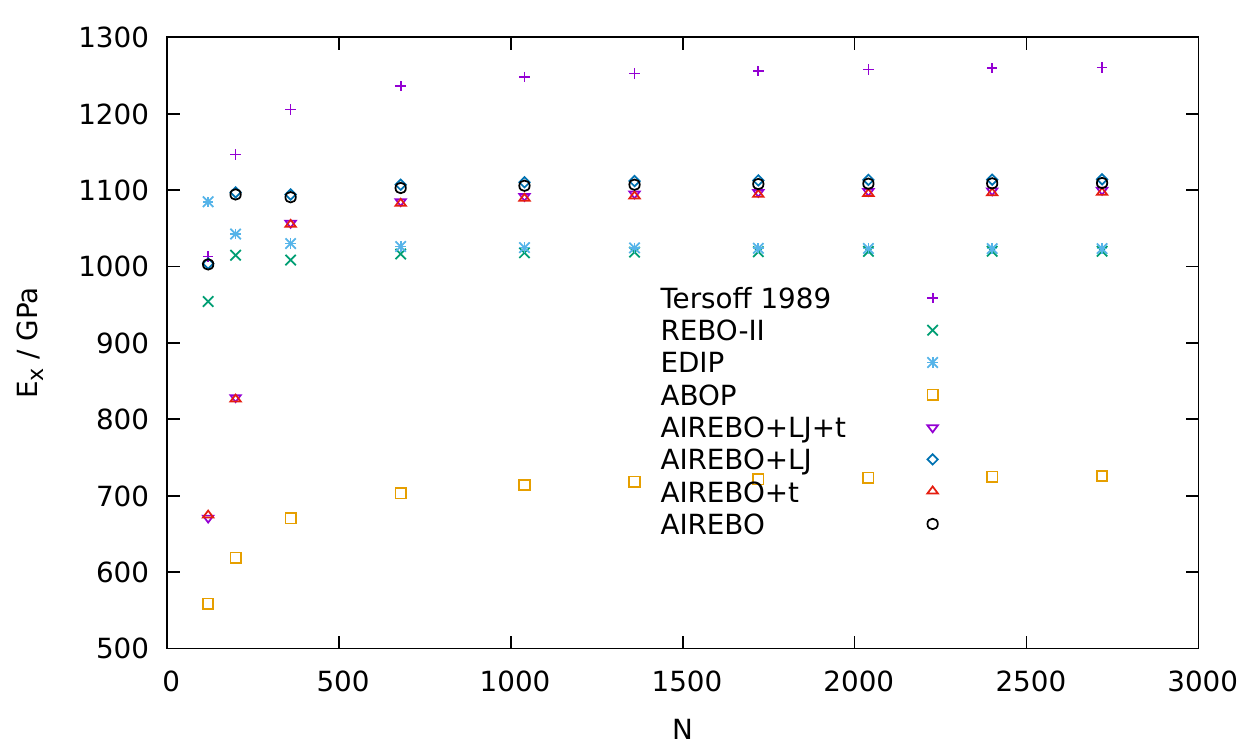}
\includegraphics*[clip,width=80mm,keepaspectratio]{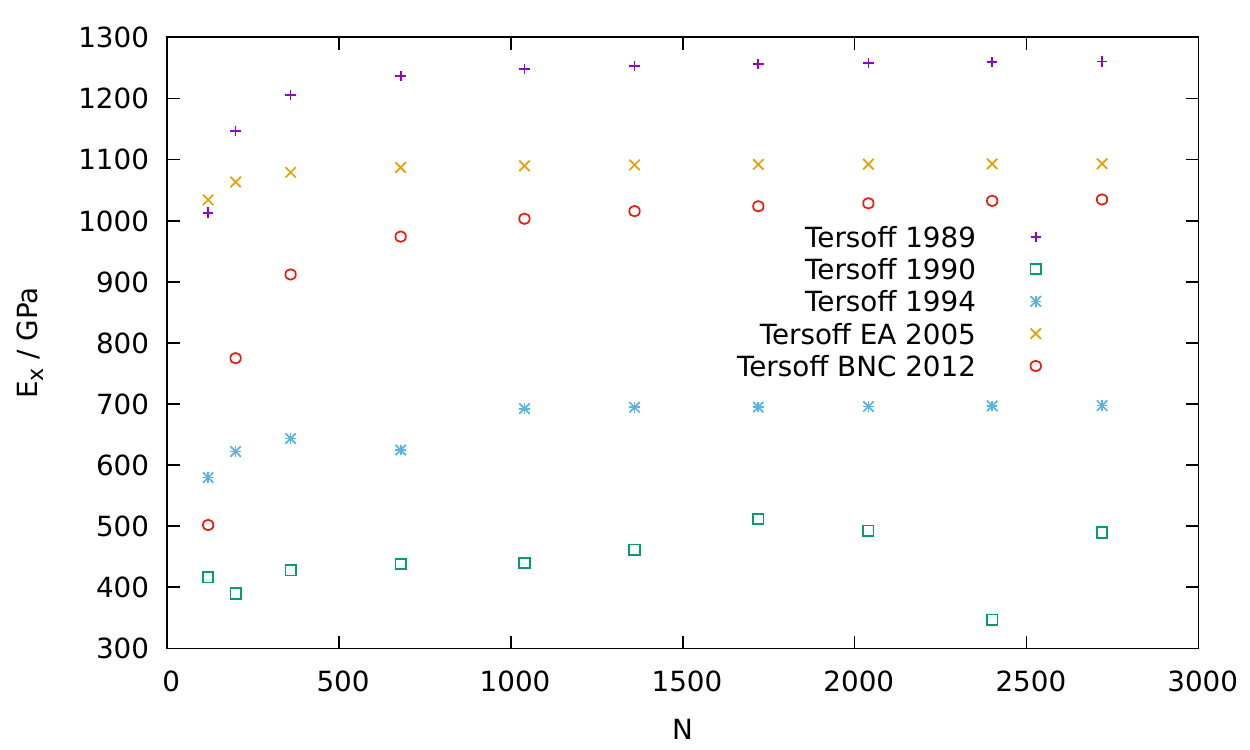}
\caption{Young's modulus of  a (20,20) CNT with
armchair geometry along the tube, taken as $x$-direction, for
various sizes and potentials.}  
\label{cmd-modulus-cnt}
\end{figure}
%===================    figure =================================

Since CNTs share the $sp^2$ structure with graphene, one would expect that
Young's moduli of single walled CNTs are very similar to that of
graphene, which is indeed the case at least for large enough
radii \cite{MTZ:CPL04,XSG:MSEA10}. For our calculations this
similarity also holds. Again, the majority of potentials
converges with $N$ against values in the range of now $1.0\dots
1.3$~TPa, see l.h.s. of \figref{cmd-modulus-cnt}. The various
investigated AIREBO potentials once more yield 
identical results. The EDIP potential comes closest to 1~TPa,
again together with REBO-II,
whereas the modulus calculated with  ABOP again stays below 0.8~TPa.

Also for the Tersoff potentials we obtain results similar to
those for graphene, compare r.h.s. of
\figref{cmd-modulus-cnt}. The large deviation for the Tersoff
potentials of 1990 and 1994 correlates again with deficiencies to
reproduce the structure. Using the version of 1994 the transverse
section of the CNT is not a circle but more a rounded square in
our simulations, whereas we could not obtain a reasonable
structure with the 1990 version at all.

For C-C- bond distances compare table~\ref{tab-cmd-modulus-1}.

%%%%%%%%%%%%%%%%%%%%%%%%%%%%%%%%%%%%%%%%%%%%%%%%%%%%%%%%%%%%%%%%%%%%%
\subsection{Diamond}
\label{sec-3-3}

The studied diamond structures had a size of about $\sqrt[3]{N}
\times\sqrt[3]{N} \times\sqrt[3]{N}$. We performed simulations up to
linear sizes of 20 atomic positions, which appears to be sufficient for the
conclusions of this paper.

%===================    figure   =================================
%\begin{figure}[ht!]
\begin{figure}
\centering
\includegraphics*[clip,width=80mm,keepaspectratio]{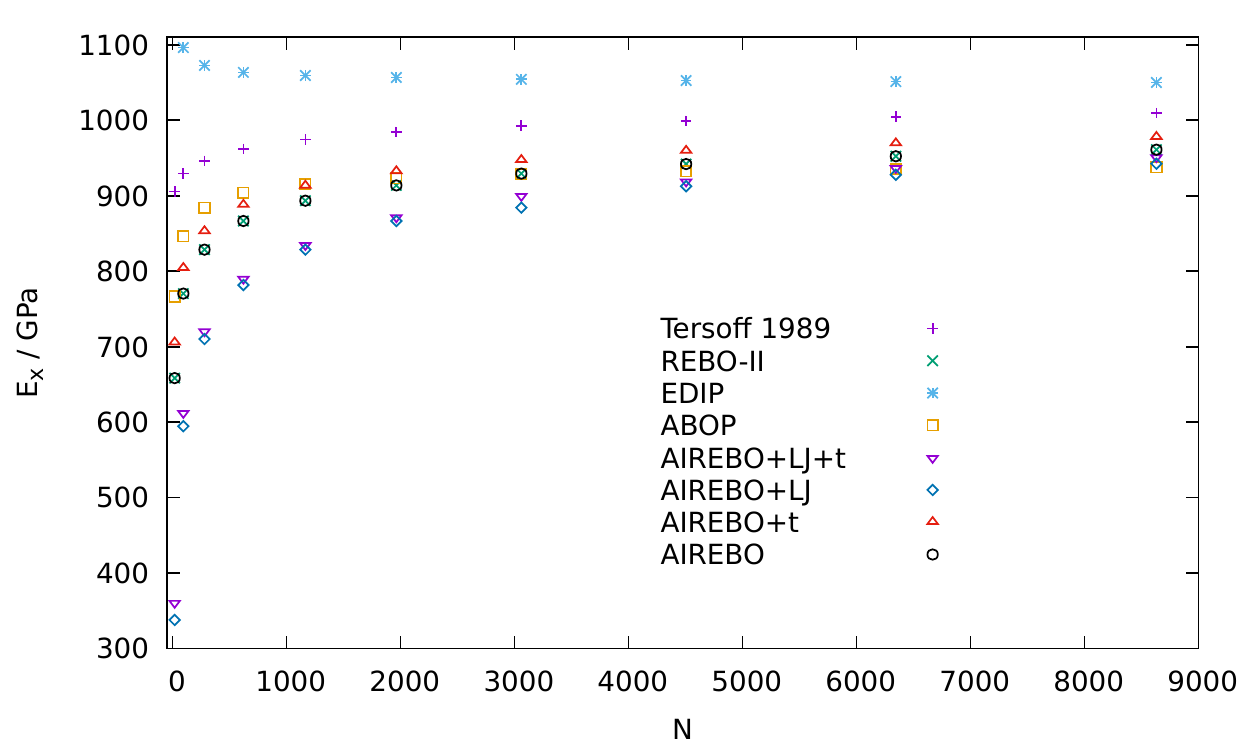}
\includegraphics*[clip,width=80mm,keepaspectratio]{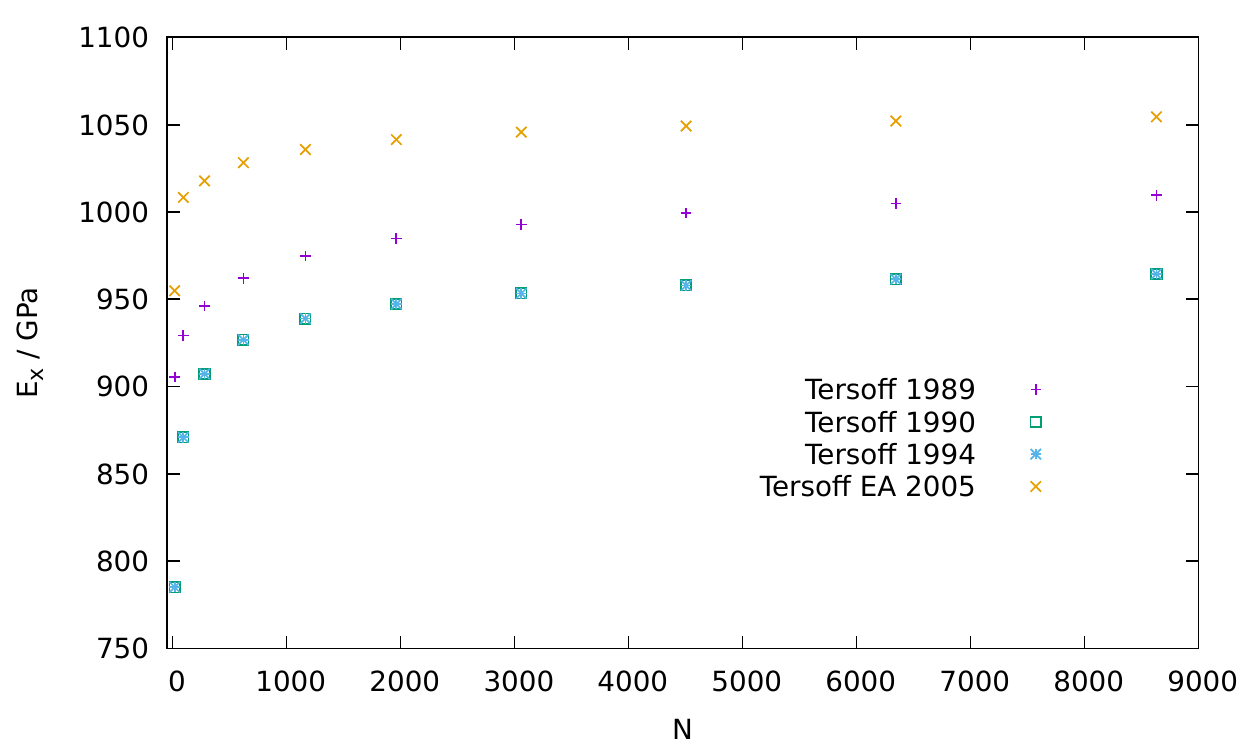}
\caption{Young's modulus of diamond taken along the
  $x$-direction shown in \figref{cmd-modulus-diamond-directions} for
  various sizes and potentials.} 
\label{cmd-modulus-diamond}
\end{figure}
%===================    figure =================================

The experimental value of Young's modulus was determined rather
early in 1940 as 
$5.5\times 10^{12}$ dynes per sq. cm ($=0.55$~TPa) \cite{Pis:PIAS40},
which we would like to cite for curiosity. Since the modulus is
direction-dependent, modern investigations yield an average of
about $\overline{E}=1.15$~TPa \cite{KlC:DRM93} with values spreading between
1.05~TPa and 1.21~TPa \cite{SpD:W94}.
In our simulations we find a
rather good overall agreement between all potentials shown on
the l.h.s. of \figref{cmd-modulus-diamond}. Although not yet
fully converged to the thermodynamic limit for the largest
calculated $N$, one clearly sees 
that all results agree with $(1\pm 0.07)$~TPa. REBO-II
does not coincide with EDIP this time, but now with AIREBO without
Lennard-Jones and torsion (AIREBO). 
Among the Tersoff potentials shown on the r.h.s. of
\figref{cmd-modulus-diamond} the 2005 parameterization yields a
similarly good result, whereas the parameterizations of 1990
and 1994 again deviate towards too small values.

%===================    figure   =================================
%\begin{figure}[ht!]
\begin{figure}
\centering
\includegraphics*[clip,width=50mm,keepaspectratio]{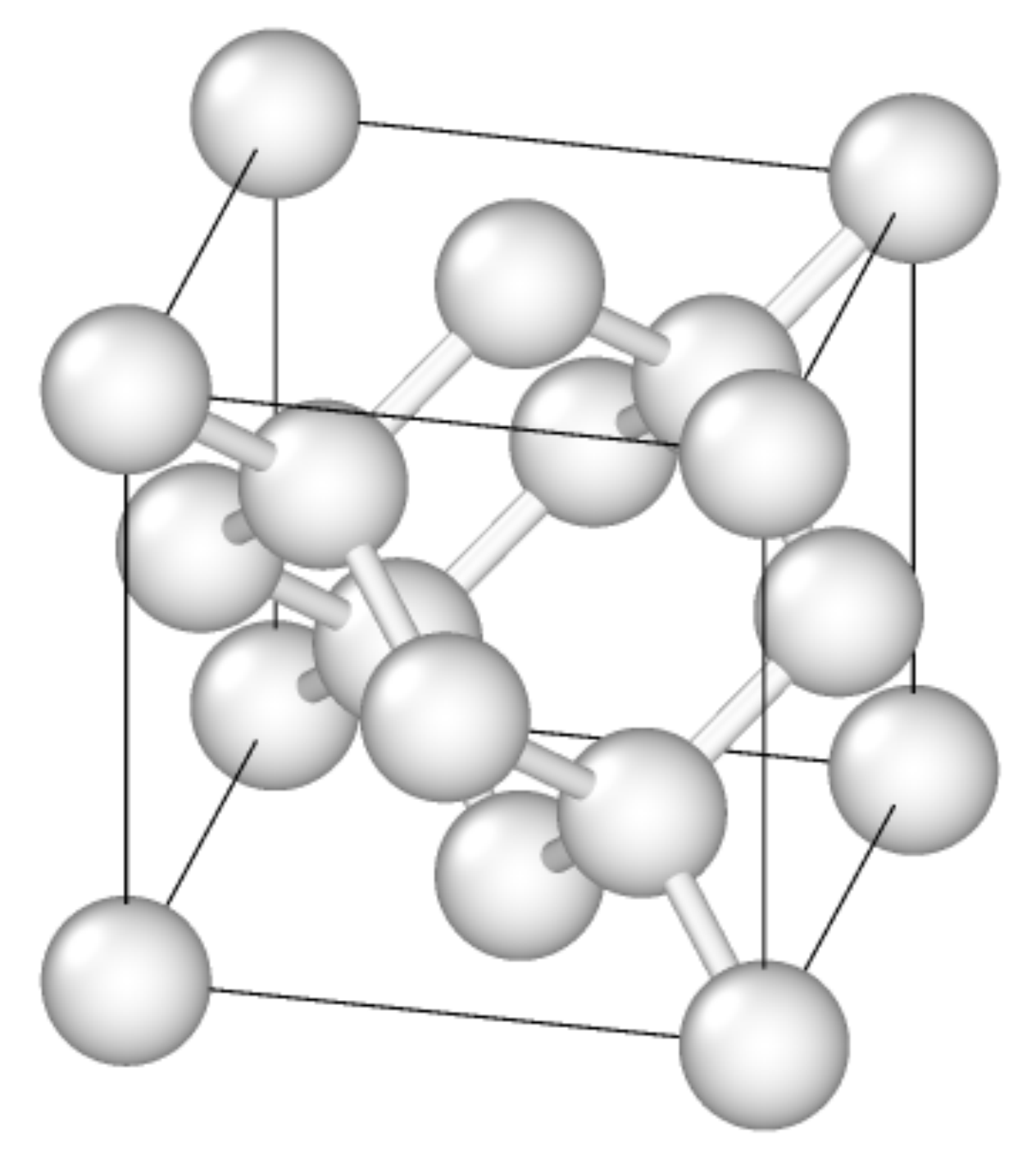}
\includegraphics*[clip,width=100mm,keepaspectratio]{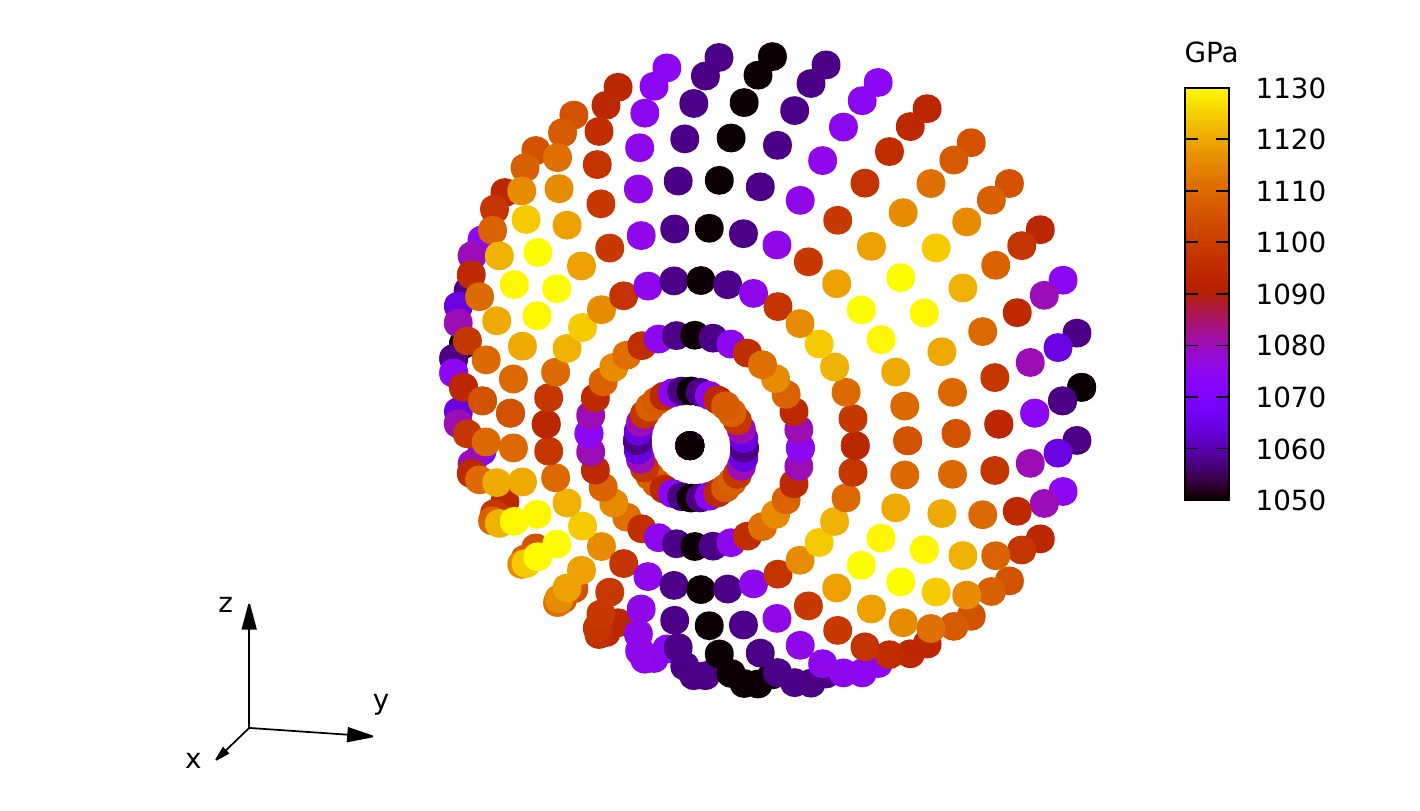}
\caption{Structure -- graphics produced with OVITO
  \cite{Stu:MSMSE10} -- and directions as well as Young's
  modulus of diamond taken in various directions 
  of the northern hemisphere around the positive $x$-direction for
  $N=8631$ and the EDIP potential.}  
\label{cmd-modulus-diamond-directions}
\end{figure}
%===================    figure =================================

The directional variation of the modulus was investigated for
the EDIP potential for the largest considered system size of
$N=8631$. As can be seen in 
\figref{cmd-modulus-diamond-directions} the potential reproduces
nicely the experimental variation of the modulus.

%===================    figure   =================================
%\begin{figure}[ht!]
\begin{figure}
\centering
\includegraphics*[clip,width=80mm,keepaspectratio]{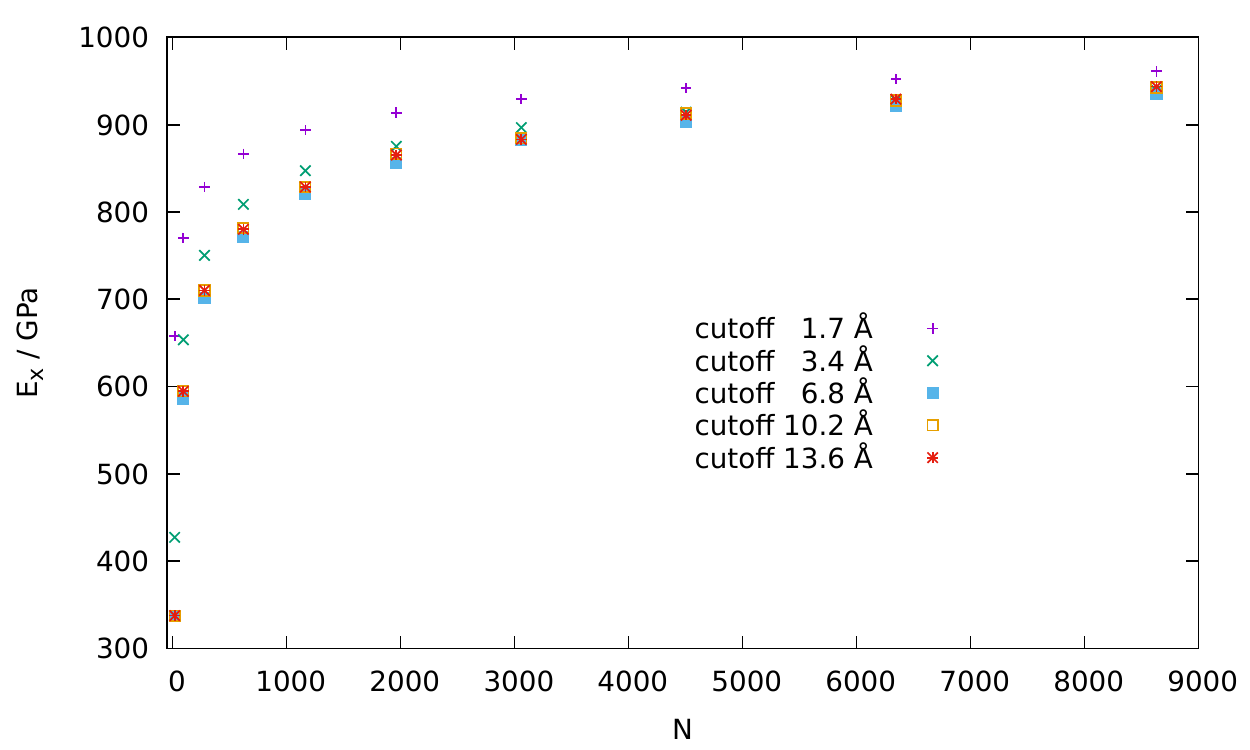}
\caption{Young's modulus of diamond taken along the
  $x$-direction shown in \figref{cmd-modulus-diamond-directions} for
  the AIREBO+LJ potential with various cutoffs.} 
\label{cmd-modulus-diamond-b}
\end{figure}
%===================    figure =================================

Besides of the importance to describe the $sp^3$-bonding
correctly, long-range interactions may play an important role in
diamond. This question is investigated in
\figref{cmd-modulus-diamond-b}, where the results derived from
the AIREBO potential with additional long-range Lennard-Jones
part, but without torsion, i.e. AIREBO+LJ, are displayed. The
result is somewhat non-intuitive, no clear dependence on the range
cutoff could be seen; all results are very close to each other
with the exception of the smallest cutoff. 

%===================    table   =================================
\begin{table}
\centering
\caption{Ground-state dimensions in \AA\ of graphene, CNT, and diamond
  for the investigated potentials. (* No proper ground state
  structure found; $\dagger$ anisotropic.)}
\label{tab-cmd-modulus-1}
\begin{tabular}{l||r|r|r}
  \hline
potential & graphene  & CNT & diamond \\
          & C-C distance & C-C distance & lattice const.\\
  \hline
EDIP \cite{Mar:PRB00}       & 1.42   & 1.42	     & 3.56\\	   
REBO-II \cite{BSH:JPCM02}       & 1.42   & 1.42	     & 3.58\\	   
ABOP \cite{ZWF:JCC15}       & 1.42   & 1.424, 1.417 $\dagger$ & 3.46\\	   
Tersoff 89 \cite{Ter:PRB89} & 1.46   & 1.46          & 3.57  \\ 
Tersoff 90 \cite{Ter:PRL90} & *      & *             & 3.56  \\ 
Tersoff 94 \cite{Ter:PRB94} & 1.55   & *             & 3.56 \\  
Tersoff BNC \cite{PhysRevB.86.115410,PhysRevB.81.205441} & 1.44   & 1.44          & -\\
Tersoff EA  \cite{PhysRevB.71.035211} & 1.48  & 1.48         & 3.57   \\
AIREBO+LJ+t \cite{STH:JCP00} & 1.40   & 1.41         & 3.58 \\
AIREBO+LJ  \cite{STH:JCP00} & 1.40   & 1.40         & 3.58  \\
AIREBO+t  \cite{STH:JCP00} & 1.40   & 1.40         & 3.58	\\  
AIREBO  \cite{STH:JCP00} & 1.40   & 1.40         & 3.58    \\
\hline
experimental & 1.42   & 1.42         & 3.567    \\
\hline
\end {tabular}
\end {table}
%===================    table   =================================

As additional information about the performance of classical
carbon potentials implemented in LAMMPS we provide ground-state
C-C distances for graphene and the CNT as well as the lattice
constant for diamond in table~\ref{tab-cmd-modulus-1}. As one
can see, not all potentials perform equally well with respect to
these characteristic distances. The EDIP potential meets all
experimental numbers. Since the Tersoff-2012 potential was
designed for  B, C, and BN-C hybrid based graphene like nano
structures, we did not use it for diamond.

%%%%%%%%%%%%%%%%%%%%%%%%%%%%%%%%%%%%%%%%%%%%%%%%%%%%%%%%%%%%%%%%%%%%%
\section{Discussion and Outlook}
\label{sec-4}

For the investigated observable (Young's modulus) and the chosen
carbon materials it turns out that Marks' improved EDIP
potential and REBO-II \cite{BSH:JPCM02} perform overall good
with slight 
differences for diamond. REBO-II reacts somewhat softer to elongations for
diamond which is very likely related to the smaller cutoff of
the potential.  
EDIP uses a longer range of 3.2~\AA, whereas REBO-II uses a
cosine cutoff between 1.7 and 2.0 \AA. This difference does not
matter for graphene and CNTs. Both potentials lack long-range
van der Waals components.

The potential ABOB \cite{ZWF:JCC15} is consistently too soft for
all three materials; for the $sp^2$ materials the deviation is
as large as 20~\%, for diamond the situation is better. 

Among the class of Tersoff potentials the LAMMPS implementations
of the parameterizations of
1990 \cite{Ter:PRL90} and 1994 \cite{Ter:PRB94} produce
untrustworthy results. Even the relatively 
simple ground state structures of graphene and CNTs turned out
to be wrong, compare also table~\ref{tab-cmd-modulus-1}.
Reference~\cite{LXS:ASS13} recommends the Tersoff potential for
modeling of diamond. In view of our results this recommendation
holds only for Tersoff 89 \cite{Ter:PRB89} and Tersoff EA
\cite{PhysRevB.71.035211}. 

For the observable studied in this paper we noticed that the
variants of AIREBO do not differ significantly.
We also noticed that although the LAMMPS documentation on AIREBO states
that ``If both of the LJ an torsional terms are turned off, it becomes
the 2nd-generation REBO potential, with a small caveat on the
spline fitting procedure mentioned below.'', our results for the
CNT and graphene differ by more than what would be compatible
``a small caveat''.

%%%%%%%%%%%%%%%%%%%%%%%%%%%%%%%%%%%%%%%%%%%%%%%%%%%%%%%%%%%%%%%%%%%%%
%% The "Acknowledgements" section can be given in all manuscripts.
%% This should be done within the ``acknowledgements'' environment,
%% which will make the correct section title.
%%%%%%%%%%%%%%%%%%%%%%%%%%%%%%%%%%%%%%%%%%%%%%%%%%%%%%%%%%%%%%%%%%%%%
\begin{acknowledgements}
We are very thankful to Prof. Nigel Marks for sharing with us
the details of his carbon-carbon potential.
\end{acknowledgements}

%\bibliography{cmd-modulus}

%%%%%%%%%%%%%%%%%%%%%%%%%%%%%%%%%%%%%%%%%%%%%%%%%%%%%%%%%%%%%%%%%%%%%
\vspace{3cm}
This article is published in full length in \textit{Beilstein
  J. Nanotechnol.}
\textbf{20??}, \textit{?}, No. ?.

%%%%%%%%%%%%%%%%%%%%%%%%%%%%%%%%%%%%%%%%%%%%%%%%%%%%%%%%%%%%%%%%%%%%%
%% That's it. Ending the document finishes the article. Happy TeXing!
%%%%%%%%%%%%%%%%%%%%%%%%%%%%%%%%%%%%%%%%%%%%%%%%%%%%%%%%%%%%%%%%%%%%%
\end{document}